\documentclass[
twocolumn,
 amsmath,amssymb,
 aps,
pra,
]{revtex4-2}

\usepackage{graphicx}
\usepackage{dcolumn}
\usepackage{bm}
\usepackage{verbatim}
\usepackage{hyperref}
\usepackage{color}

\usepackage{enumitem}

\begin{document}

\title{Dissipative dynamics of an impurity with spin-orbit coupling}

\author{Areg Ghazaryan, Alberto Cappellaro, Mikhail Lemeshko, Artem G. Volosniev}
 \affiliation{Institute of Science and Technology Austria (ISTA), am Campus 1, 3400 Klosterneuburg, Austria}
\date{\today}

\begin{abstract}
Brownian motion of a mobile impurity in a bath is affected by spin-orbit coupling~(SOC). Here, we discuss a Caldeira-Leggett-type model that can be used to propose and interpret quantum simulators of this problem in cold Bose gases. First, we derive a master equation that describes the model and explore it in a one-dimensional (1D) setting. To validate the standard assumptions needed for our derivation, we analyze available experimental data without SOC; as a byproduct, this  analysis suggests that the quench dynamics of the impurity is beyond the 1D Bose-polaron approach at temperatures currently accessible in a cold-atom laboratory -- motion of the impurity is mainly driven by dissipation. For systems with SOC, we demonstrate that 1D spin-orbit coupling can be `gauged out' even in the presence of dissipation -- the information about SOC is incorporated in the initial conditions. Observables sensitive to this information (such as spin densities) can be used to study formation of steady spin
polarization domains during quench dynamics.
\end{abstract}

\maketitle

Dissipation of energy occurs naturally when a particle with finite momentum moves through a medium. This phenomenon is typically studied assuming that the momentum of the particle is decoupled from its spin degree of freedom. This is however not the case for many condensed matter systems with strong spin-orbit coupling (SOC). In particular, for
externally driven setups with non-trivial topological character, such as  bosonic Kitaev-Majorana chains~\cite{mcdonald2018phase,flynn2021topology}, Majorana wires~\cite{liu2017role,huang2019dissipative}, as well as systems featuring
optical spin-Hall effect~\cite{kavokin2005optical,leyder2007observation}. 
SOC is also key for explaining transport of electrons through a layer of chiral molecules~\cite{Naaman2019,Evers2022}. To understand equilibration processes in these systems and promote their use in technologies, studies of dissipative dynamics with SOC are needed. Cold-atom-based quantum simulators provide a natural testbed for such studies~\cite{LangenReview06,Polkovnikov2011} -- they complement the existing research of out-of-equilibrium time evolution, see, e.g., \cite{DAlessio2016,Lewis2019,Abanin2019}, and SOC engineering using laser fields~\cite{Dalibard2011,Galitski2013}.

To enjoy the potential of quantum simulators, one requires theoretical models that can be used to propose new experiments and analyze the existing data~\cite{Altman2021}. In this work, we present one such model designed to study an impurity with SOC (see also recent Ref.~\cite{Takahashi2021} for a discussion of a relevant Langevin-type equation).  The impurity is in contact with the bath that we model as a collection of harmonic oscillators. Using the Born and Markov approximations, we derive a master equation, which extends the result of Caldeira and Leggett~\cite{CALDEIRA1983587, breuer2002theory}  to a spin-orbit-coupled impurity. 
To illustrate this equation, we focus on one-dimensional (1D) systems. First, we test it using the experimental data of Ref.~\cite{Catani2012}, whose full theoretical understanding is lacking, see, e.g., Ref.~\cite{Grusdt_2017}. We find that the Caldeira-Leggett model contains all ingredients to describe the observed breathing dynamics of the impurity assuming that the initial condition is the (only) tunable parameter. The calculations are {\it analytical}, which simplifies the analysis and allows us to gain insight into the system: relevant time scales, short- and long-time dynamics.
Finally, we explore the dynamics of the system with SOC. Without magnetic fields, the 1D SOC can be gauged out so that the system can be described using the Caldeira-Leggett equation with SOC-dependent initial conditions. We present observables that are sensitive to these initial conditions and can be used to study the effect of SOC on time dynamics, for example, formation of regions with steady spin polarization.   
Our findings provide a convenient theoretical model that can be used to propose and benchmark quantum simulators of dissipative dynamics with SOC.

\textit{The particle-bath Hamiltonian}.
The Hamiltonian of the system is given by
$H_{\text{tot}} = H_S + H_B + H_C$. The three terms account for, correspondingly, the (quantum) impurity, the harmonic bath and the bath-impurity coupling. We assume that 
$H_S$ has the form
\begin{equation}
H_S  = \frac{\mathbf{p}^2}{2m} 
+ V_{SO}(\mathbf{p},\pmb{\sigma})
+ \mathcal{V}_{\text{ext}}(\mathbf{q})
+\mathbf{q}^2 \sum_{j=1}^N \frac{c_j^2}{2m_j\omega_j^2},
\label{hamiltonian system}
\end{equation}
where $m$ and $\mathbf{q}$ are the mass and the position of the impurity, respectively. $\mathcal{V}_{\text{ext}}$ is an external potential. $V_{SO}(\mathbf{p},\pmb{\sigma})$ is the potential that describes SOC;
it depends on the Pauli vector, $\pmb{\sigma}$, and the momentum of the impurity, $\mathbf{p}$. A particular form of $V_{SO}$ is specified below, see also the Supplementary Material. 
The last term in Eq.~\eqref{hamiltonian system} is a standard harmonic counterterm, which makes $H_{\mathrm{tot}}$ translationally invariant for $\mathcal{V}_{\text{ext}}=0$~\cite{Caldeira2014}.
The parameters $\omega_j$ and $m_j$ are taken from the bath Hamiltonian, $H_B =\sum_j [\mathbf{p}_j^2/(2m_j) +\frac{1}{2}m_j\omega_j^2\mathbf{x}_j^2]$; $c_j$ defines the strength of the bath-impurity interaction $H_C = -\mathbf{q} \cdot \sum_j c_j \mathbf{x}_j$. [For microscopic derivations that validate the form of $H_B$ and $H_C$ for weakly interacting Bose gases and Luttinger liquids, see correspondingly Refs.~\cite{Lampo2017} and \cite{recati-2005}.] To summarize, we consider a single particle (impurity) linearly coupled to an environment
made of non-interacting harmonic oscillators using the standard procedure~\cite{CALDEIRA1983374,CALDEIRA1983587}, briefly outlined below; this well-studied problem is extended here by subjecting the impurity to SOC. 

Before analyzing $H_{\mathrm{tot}}$, we remark that there are a number of theoretical methods ~\cite{Fermi_polaron_review,Fermi_polaron_theory_review,Bose_polaron_review,Scazza2022,Schmidt2018} that can be used for interpreting experiments with impurities in Fermi~\cite{Schirotzek2009,Koschorreck2012,Cetina2016,Scazza2017} and Bose gases~\cite{Spethmann2012,Catani2012,Hu2016,Jorgensen2016,Yan_2020,Skou2021}. Time evolution of an impurity in a Bose gas -- the focus of this work -- has been studied using variational wave functions, $T$-matrix approximations and exact solutions in 3D at zero~\cite{Volosniev2015,Shchadilova2016,Drescher2019} and finite temperatures~\cite{Dzsotjan2020}.  Many more methods exist to address the 1D world. For example, experimentally relevant trapped systems can be studied using numerically exact approaches~\cite{Peotta2013,Mistakidis2019}, for review see Ref.~\cite{Mistakidis2022}. In cases when those methods do not work (e.g., large energy exchange or high temperature), it has been suggested to connect a cold-atom impurity to quantum Brownian motion~\cite{Massignan2015,Lampo2017,Lampo2018,Nielsen2019}. Our work provides an example when this idea leads to an accurate description of experimental data, setting the stage for testing assumptions behind theoretical models of relaxation~\cite{breuer2002theory,Caldeira2014} in a cold-atom laboratory. 

\textit{Born-Markov master equation with SOC}. 
Time evolution of the impurity-bath ensemble, defined by 
$H_{\text{tot}}$, obeys the Von-Neumann equation: 
$i\hbar \dot{\rho}_{\text{tot}} = [H_{\text{tot}},\rho_{\text{tot}}]$. To extract dynamics of the impurity from $\rho_{\text{tot}}$, we rely on the Born-Markov approximation
\cite{breuer2002theory,lebellac,gardiner2004quantum}, which leads to the equation for the (reduced) density matrix that describes the impurity, $\rho_S$:
\begin{equation}
\begin{aligned}
\frac{d\rho_S}{dt} & = - \frac{i}{\hbar}\big[H_S,\rho_S\big] 
- \frac{1}{\hbar^2} \int_0^{+\infty} ds\, \mathcal{C}(s)\bigg[
\mathbf{q},\big[\mathbf{Q}(-s),\rho_S\big]
\bigg] \\
& \quad+ \frac{i}{\hbar^2}\int_0^{+\infty} ds\, \chi(s)\bigg[\mathbf{q},\lbrace \mathbf{Q}(-s), 
\rho_S\rbrace\bigg]\;.
\end{aligned}
\label{born-markov master equation}
\end{equation} 
We write Eq.~(\ref{born-markov master equation}) in the form standard for a Brownian particle; the contribution of SOC is conveniently hidden in $\mathbf{Q}(t)$, which is defined as
\begin{equation}
\mathbf{Q}(t) = \frac{i}{\hbar} [H_S,\mathbf{q}] = \mathbf{q}
-\bigg(\frac{\mathbf{p}}{m} +\mathbf{v}_{SO}(\pmb{\sigma})\bigg)\,t \,,
\label{position operator interaction}
\end{equation}
where $\mathbf{v}_{SO} = \partial_{\mathbf{p}} V_{SO}$ is the contribution to the `velocity' of the particle due to SOC.
Equation~(\ref{born-markov master equation}) contains the bath 
autocorrelation functions $\mathcal{C}(t)$ and $\chi(t)$ in 
Eq. \eqref{born-markov master equation}
\begin{equation}
\begin{aligned}
\mathcal{C}(t) &= 
 \hbar \int_0^{+\infty} d\omega\,J(\omega) \coth\bigg(\frac{\beta\hbar\omega}{2}\bigg) \cos(\omega t), \\
\chi(t)  & = 
\hbar\int_0^{+\infty} d\omega\, J(\omega) \sin(\omega t)\;,
\end{aligned}
\label{autocorrelations}
\end{equation}
where $\beta=1/{k_B T}$ ($T$ for temperature, and $k_B$ is the Boltzmann constant).
These functions assume that all relevant microscopic information is encoded in the spectral function 
$J(\omega)$, formally defined as $J(\omega) = \sum_{j} c_j^2 \delta (\omega-\omega_j)/(2m_j\omega_j)$. 
We choose 
$J(\omega) = (2m\gamma/\pi) \, \omega \, \Omega_c^2/(\Omega_c^2 + \omega^2)$,
recovering Ohmic dissipation at $\omega \rightarrow 0$; the phenomenological
parameter $\Omega_c$ defines the high-frequency behavior.
The Ohmic spectral density is a standard choice in mesoscopic 
\cite{breuer2002theory,leggett-1987,lehur-2016} and in cold-atom physics 
\cite{recati-2005,orth-2008,dalidovich-2009}.
We employ it here because it leads to a local-in-time damping that agrees with the experimental data used below to validate the model (see also Refs.~\cite{recati-2005,popov-book} for additional details about Ohmic dissipation in 1D
based upon long-wavelength approximations for superfluids).
Super-Ohmic dissipation whose relevance 
for Bose polarons is highlighted in Refs.~\cite{Lampo2017,Lampo2018} leads to strong memory effects (non-local-in-time damping), thus, we do not consider it here.

\begin{figure*}
  \includegraphics[width=\textwidth]{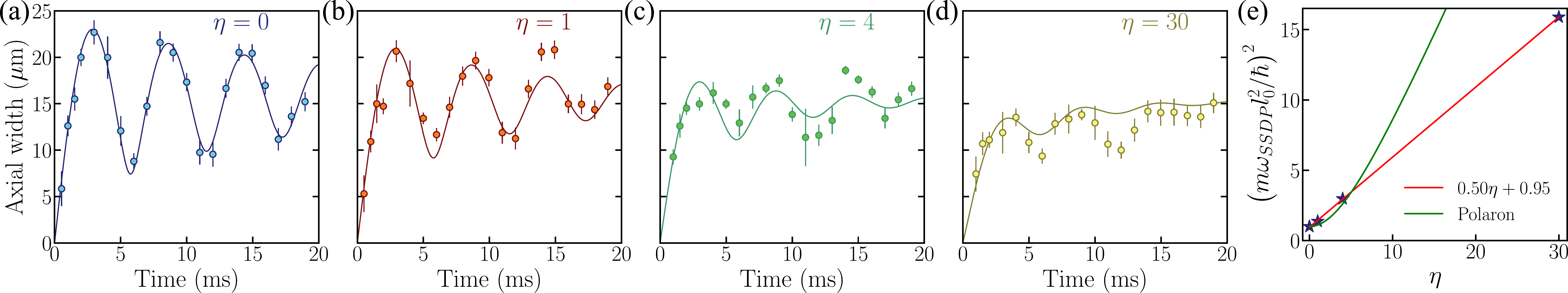}
  \caption{(a)-(d) The width of the impurity (potassium) cloud $\bar y$ as a function of time for different values of the parameter $\eta$. The dots with error bars show the experimental data of Ref.~\cite{Catani2012}. The curves are the fits to Eq.~(\ref{master position_main}). Panel (e) shows the values of $l_0$ used in the fit as a function of $\eta$. The panel also shows a linear fit to these values  (red line). The green curve shows the effective mass of the polaron calculated using the analytical methods outlined in Refs.~\cite{Mistakidis2018,  Jager2020} (no fitting parameters).}
  \label{fig:fit_TFfixed}
\end{figure*}

Using Eqs.~\eqref{position operator interaction} and~\eqref{autocorrelations}, we derive the master equation
\begin{align}
\frac{d\rho_S}{dt} &= - \frac{i}{\hbar}[{H}_S,\rho_S] - \frac{i\gamma}{\hbar}\,\big[
\mathbf{q},\big\lbrace \mathbf{p},\rho_S\big\rbrace \big]
- \frac{2 m\gamma}{\beta\hbar^2} \big[\mathbf{q},\big[ \mathbf{q},\rho_S\big]\big]\nonumber \\
& \qquad - im\gamma\big[\mathbf{q},\big\lbrace
{\mathbf{v}}_{SO}, \rho_S
\big\rbrace \big], 
\label{modified master equation}
\end{align}
where $\gamma$ defines the strength of dissipation.
The frequency integrals leading
to Eq.~\eqref{modified master equation} are discussed in the Supplementary Material. 
As expected, dissipative dynamics is affected by SOC, see the last term in Eq.~(\ref{modified master equation}).
Finally, a proper Lindblad form for Eq.~\eqref{modified master equation} can be achieved by
adding a \textit{minimally invasive} term: $-\gamma\beta\left[\mathbf{p}\left[\mathbf{p},\rho_S\right]\right]/(8m)$~\cite{breuer2002theory,ferialdi2017dissipation}; we employ this term in our calculations.

To illustrate the master equation, we choose to consider a 1D setting parameterized by the coordinate $y$. Without loss of generality, we write the SOC term as $V_{SO}=\alpha\sigma_xp_y$.
In this case, the master equation reads as 
\begin{equation}
\begin{aligned}
\frac{d{\rho}}{dt}=\frac{d{\rho}}{dt}\bigg|_{\alpha=0}-\alpha \mathcal{F}[\rho],
\end{aligned}
\label{master position_main}
\end{equation}
where $\mathcal{F}[\rho]=\sigma_x\partial_y{\rho}+\partial_{y^\prime}{\rho}\sigma_x +\frac{im\gamma}{\hbar} (y-y')\big(\sigma_x{\rho} + {\rho}\sigma_x \big)$ with
${\rho}\equiv\langle y\left|\rho_S\right|y^\prime\rangle$, and $\frac{d{\rho}}{dt}|_{\alpha=0}$ describes time evolution of the system without SOC (see the Supplementary Material). 
The effect of SOC is encoded in $\alpha \mathcal{F}[\rho]$.

While the technical details leading to Eq. \eqref{born-markov master equation}
are presented in the Supplementary Material, we recall here the standard assumptions behind the
Born-Markov approximation. First, the impurity-bath density matrix is separable throughout time evolution,
such that $\rho_{\text{tot}}(t) \simeq \rho_S(t) \otimes \rho_B(t)$. Second,
the bath is not affected by the impurity motion, namely, $\rho_B(t) \simeq 
\rho^{\text{eq}}_B$. This assumption is natural if the decay of bath correlations
has the fastest timescale $\tau_B$; it implies that dynamical features 
$\sim \tau_B$ are not resolved by our approach~\cite{breuer2002theory,lebellac}. To validate  these approximations, we shall demonstrate that the master equation is capable of describing experimental data of Ref.~\cite{Catani2012} that provide a benchmark point for us at $\alpha=0$.

{\it Dynamics without SOC}. 
First, we briefly outline the main features and findings of the experiment of Ref.~\cite{Catani2012}. In that work, a potassium atom was used to model an impurity in a gas of rubidium atoms. At $t=0$, the impurity was trapped in a tight trap created by a species-selective dipole potential (SSDP) with $\omega_{\mathrm{SSDP}}/(2\pi)=1$kHz. At $t>0$, the dynamics was initiated by an abrupt removal of the SSDP; the impurity was still confined by a shallow parabolic potential, i.e., $\mathcal{V}_\mathrm{ext}\left(y\right)=\hbar^2y^2/2ml^4$, where $l=\sqrt{\hbar/m\omega}$ and $\omega=(87\times 2\pi)$Hz is the frequency of the oscillator.
The experiment recorded the size of the impurity cloud $\bar y=\sqrt{\langle y^2 \rangle}$, and found that it can be fit using the expression
\begin{equation}
    \bar y=\bar y_0+\mathcal{A}_1 t - \mathcal{A}_2 e^{-\Gamma \Omega t}\cos[\sqrt{1-\Gamma^2}\Omega(t-t_0)],
    \label{eq:fit_experiment}
\end{equation}
where $\mathcal{A}_1, \mathcal{A}_2, \Omega, \Gamma, \bar y_0, t_0$ are fitting parameters.
The key experimental findings of Ref.~\cite{Catani2012} were: (a) $\Omega$ (almost) does not depend on the impurity-boson interaction parametrized by $\eta$; (b)  by increasing $\eta$ one decreases the amplitude of the first oscillation; 
(c) at long times $\bar y$ equilibrates to about the same value, which is independent of $\eta$. Point (b) was attributed to renormalization of the mass of the impurity, i.e., to a polaron formation~\cite{Alexandrov2009}. However, this posed several theoretical problems. In particular, the breathing frequency of the polaron cloud should depend on $\eta$, which contradicts  observation (a), see also discussions in Ref.~\cite{Catani2012,Johnson_2012,Grusdt_2017,Mistakidis2018}. Our results below suggest that one can understand the data of Ref.~\cite{Catani2012} from the perspective of dissipative dynamics.

 Equation~\eqref{master position_main} leads naturally to the dynamics observed in the experiment. To show this, we assume that the initial density matrix of the impurity corresponds to a Gaussian wave packet
\begin{equation}
\rho\left(y,y^\prime,t=0\right)=e^{-\frac{y^2+{y^\prime}^2}{2l_0^2}}/(\sqrt{\pi}l_0),
\label{eq:initial_condition_alpha=0}
\end{equation}
where $l_0$ is the parameter that determines the initial distribution of the impurity momenta; Eq.~(\ref{eq:initial_condition_alpha=0}) is standard for particles whose initial state is not precisely known.
We calculate the time dynamics for this initial condition analytically using the method of characteristics (see the Supplementary Material and Ref.~\cite{roy1999exact}), which
discovers characteristic curves where the master equation can be written as a family of ordinary differential equations~\cite{Zachmanoglou1976}.
The computed functional dependence resembles Eq.~\eqref{eq:fit_experiment} with $\Gamma \Omega= 2\gamma$ (see the Supplementary Material). Note that our calculations have only two phenomenological parameters $\gamma$ and $l_0$. All other parameters that appear in Eq.~(\ref{eq:fit_experiment}), i.e., $\mathcal{A}_1, \mathcal{A}_2,\bar y_0, t_0$ and $\Omega$, can be extracted from our results. For example, $\Omega\simeq 2\omega$ as in the experiment.

We present analytical results of the master equation together with the experimental data in Fig.~\ref{fig:fit_TFfixed}. The value of $\gamma$ is restricted to be within the errorbars of the experimentally measured value of $\Gamma\Omega$ (so that $\gamma\sim 40$Hz)~\footnote{We checked that $\gamma$ could be even fixed to the central value reported in the experiment, without affecting much the quality of the fit.}.  The temperature is set to the value reported in the experiment, i.e., $T= 350$nK~\footnote{The accuracy of the fit could be slightly improved by allowing the temperature to vary within the experimental error bars. We do not do it here to avoid having an additional fitting parameter.}. The quality of the fits in Fig.~\ref{fig:fit_TFfixed} is comparable to what can be obtained with Eq.~\eqref{eq:fit_experiment}, allowing us to conclude that the master equation provides a valuable tool for analyzing these data, and cold-atom systems in general.  

Let us briefly discuss implications of our results for interpretation of the experiment of Ref.~\cite{Catani2012}. First, the weak dependence of $\Omega$ on $\eta$ is natural in our model: the renormalization of the frequency is given by $\omega_{\mathrm{eff}}\simeq \omega(1-\gamma^2/(2\omega^2))$, where $\gamma/\omega$ is a small parameter as in the experiment. Second, the parameter $\bar y$ for $t\to \infty$ is independent of $\gamma$ assuming that the thermal de Broglie wavelength is small. Indeed, in this case, we derive $\bar y \simeq \sqrt{k_B T/(\hbar \omega)}l\approx15.42\, \mathrm{\mu m}$ in agreement with the measurement.

The amplitude of the first oscillation is determined in our analysis by the initial condition, i.e., $l_0$. To explain values of $l_0$ obtained in our fit, one can speculate that the impurity forms a polaron state at $t<0$ and that at $t>0$ the dynamics is dominated by finite temperature effects. In this picture, the mass of the impurity is renormalized (i.e., $m\to m_p$) only before the quench dynamics~\footnote{In this interpretation, the initial state is given by the polaron described by the Hamiltonian $-\frac{\hbar^2}{2m_p}\frac{\partial^2}{\partial y^2}+\frac{m\omega_{\text{SSDP}}^2y^2}{2}$. At $t>0$, the polaron is destroyed, which can be due to the anomalous behavior of the residue~\cite{Pastukhov2017} or a highly-non-equilibrium nature of the problem (cf.~Ref.~\cite{Koutentakis2022}). To show the existence of the polaron at $t=0$, one needs to consider ground-state properties of an impurity in a tight trap -- hence with a high kinetic energy -- which is beyond the scope of the present work.}; this  might explain why theoretical calculations can produce features of the amplitude but not of the frequency~\cite{Johnson_2012,Catani2012,Grusdt_2017}.
Renormalization of the mass implies that the energy scale at $t=0$ is given by $\hbar\omega_{\text{SSDP}}\sqrt{m/m_p}$, which is incorporated into Eq.~\eqref{eq:initial_condition_alpha=0} if
$l_0^2\sim \hbar/(m\omega_{\text{SSDP}})\sqrt{m_p/m}$~\footnote{Here, we calculate the expectation value of the kinetic energy for a free particle described by the Gaussian wave packet: $-\frac{\hbar^2}{2m}\langle\frac{\partial^2}{\partial y^2}\rangle$, and relate it to $\hbar\omega_{\text{SSDP}}\sqrt{m/m_p}$, which is the typical energy scale of the `polaron' Hamiltonian: $-\frac{\hbar^2}{2m_p}\frac{\partial^2}{\partial y^2}+\frac{m\omega_{\text{SSDP}}^2y^2}{2}$.}. This expression agrees qualitatively with the outcome of our fit, see Fig.~\ref{fig:fit_TFfixed}~(e). The linear increase of $(m\omega_{\text{SSDP}}l_0^2/\hbar)^2$ however quantitatively disagrees with calculations of the effective mass~\cite{Parisi2017,Grusdt_2017,PANOCHKO2019,Mistakidis2018}. The agreement improves if we disregard the point with $\eta=30$, which (as suggested in Ref.~\cite{Catani2012}) is already beyond a simple one-dimensional treatment~\footnote{One expects that large values of $\eta$ require a beyond-linear-coupling treatment of impurity-bath interactions~\cite{Grusdt_2017,Kain2018}, which is beyond the scope of the present paper.}.
 In any case, a further analysis of the experimental data (beyond the scope of this paper) is needed in light of our results.

Finally, we note that the inhomogeneity of the bath as well as non-Markovian physics do not appear to be important to describe dynamics discussed here. This stands in contrast to what is known about properties of the corresponding ground state~\cite{Dehkharghani2015} and low-energy dynamics~\cite{Johnson_2012,Schecter2016,Mistakidis2018}, and results from a high temperature and a large energy ($\sim 1/l_0^2$) associated with the initial impurity state.  

\begin{figure}[t]
  \includegraphics[width=\columnwidth]{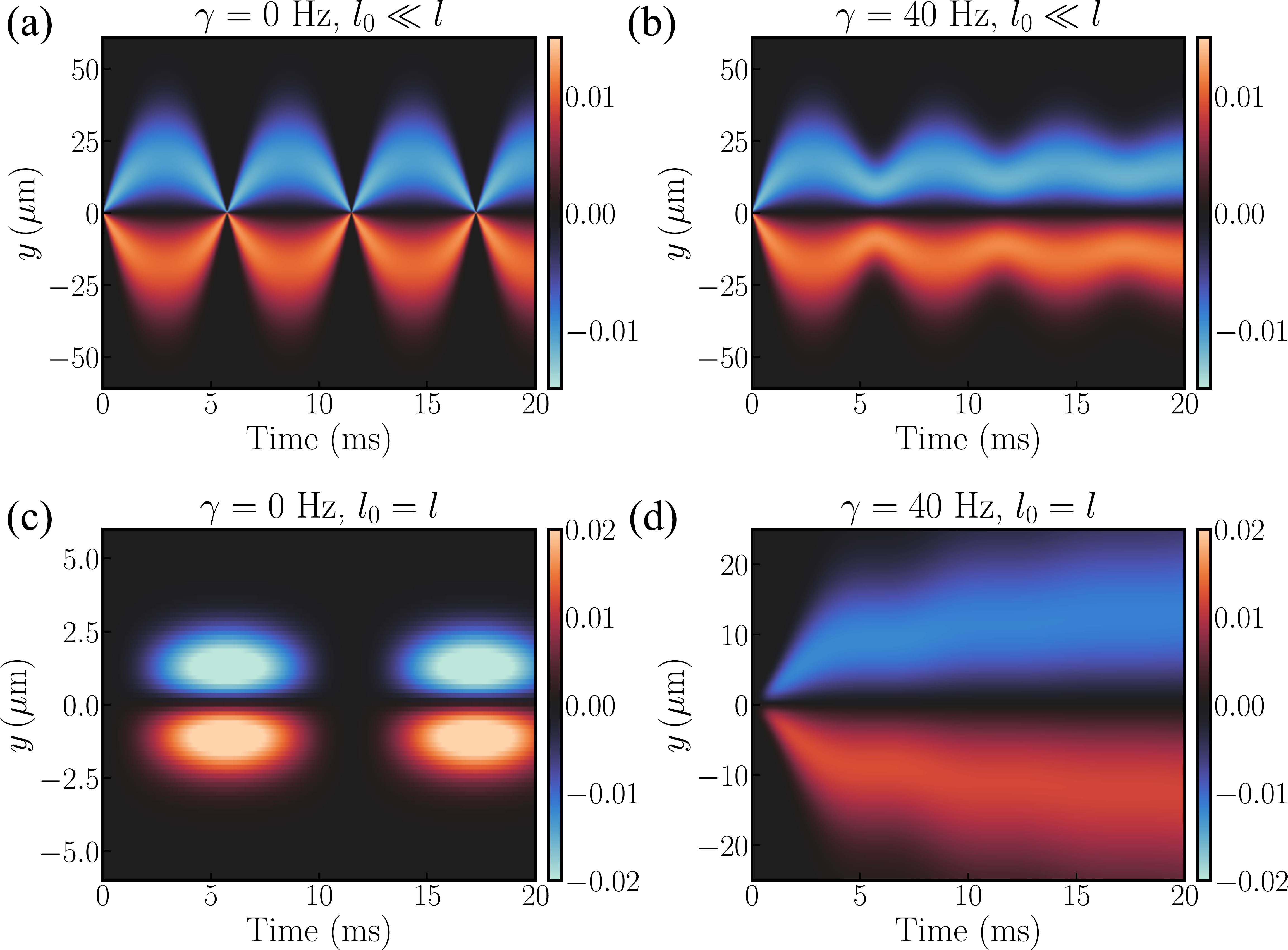}
  \caption{The spin polarization along the $y$ axis as a function of position and time in the presence of SOC. The SOC amplitude is $\alpha=40\,\mathrm{Hz\cdot\mu m}$. $l_0$ for (a,b) case corresponds to $\omega_0/2\pi=30\,\mathrm{kHz}$, while $l$ is given by $\omega/2\pi=87\,\mathrm{Hz}$. All other parameters are as in Ref.~\cite{Catani2012}, in particular $T$=350nK.}
  \label{fig:fig2}
\end{figure}

{\it Dynamics with SOC}. We use the experimental protocol of Ref.~\cite{Catani2012} also to illustrate the master equation with SOC. The peculiarity of 1D is that the $\alpha$-dependent term can be gauged out from Eq.~(\ref{master position_main}) via the transformation (in the position space) $\rho=e^{-\frac{i m\alpha \sigma_x y}{\hbar}}f e^{\frac{i m\alpha \sigma_x y^\prime}{\hbar}}$. The function $f\left(y,y^\prime,t\right)$ then satisfies the standard Caldeira-Leggett equation and can be solved exactly as without SOC (see the Supplementary Material).  Note that the equation for $f$ is spin-independent. The initial condition of the problem, $\rho\left(y,y^\prime,t=0\right)$, defines the full spin structure of the problem and time dependence of 
spin observables, as we illustrate below for $\bar \sigma_y(y,t)\equiv\mathrm{Tr}_{\mathrm{spin}}(\sigma_y)$. 

For the sake of discussion, as the initial condition we consider the state that is spin-polarized along the $z$-axis:
\begin{equation}
\rho\left(y,y^\prime,0\right)=\frac{1}{2\sqrt{\pi}l_0} e^{-\frac{y^2+{y^\prime}^2}{2l_0^2}} |\uparrow\rangle \langle\uparrow|;
\label{eqn:InitialSpatial}
\end{equation}
 other parameters of the system are taken from Ref.~\cite{Catani2012}. We use $\gamma=40$ Hz, which was typical in that experiment. The strength of SOC, $\alpha$, can be tuned in cold-atom set-ups, see, e.g.,~\cite{Struck2014,Luo2016,Shteynas2019}. We assume that $\alpha \bar y/(\omega l^2)\ll1$ to demonstrate that even weak SOC can lead to an observable effect in dynamics.

Time evolution of $\bar \sigma_y(y,t)$ is shown in Fig.~\ref{fig:fig2}. Note that 
Eq.~(\ref{eqn:InitialSpatial}) is not an eigenstate of the system with SOC -- dynamics occurs even without a change of the trap, i.e., $l_0=l$ [see Figs.~\ref{fig:fig2}~(c) and~(d)]. 
Without dissipation ($\gamma=0$), we observe oscillation of the spin density with $\bar \sigma_y>0$ for $y>0$ and $\bar \sigma_y<0$ for $y<0$ [see Figs.~\ref{fig:fig2}~(a) and~(c)]. This effect is solely due to SOC, and can be easily understood from a one-body Schr{\"o}dinger equation. Effects of temperature and dissipation are most visible in Fig.~\ref{fig:fig2}~(d): the impurity is heated up by the presence of the bath, which creates regions with steady spin polarization along the $y$ direction. Spatial extension of these regions is determined by the temperature; the time scale for their formation is given by $1/\gamma$ (similarly to the dynamics without SOC, see Fig.~\ref{fig:fit_TFfixed}). This effect can be observed in cold-atom systems by analyzing populations of the involved hyperfine states.

\begin{figure}[t]
  \includegraphics[width=\columnwidth]{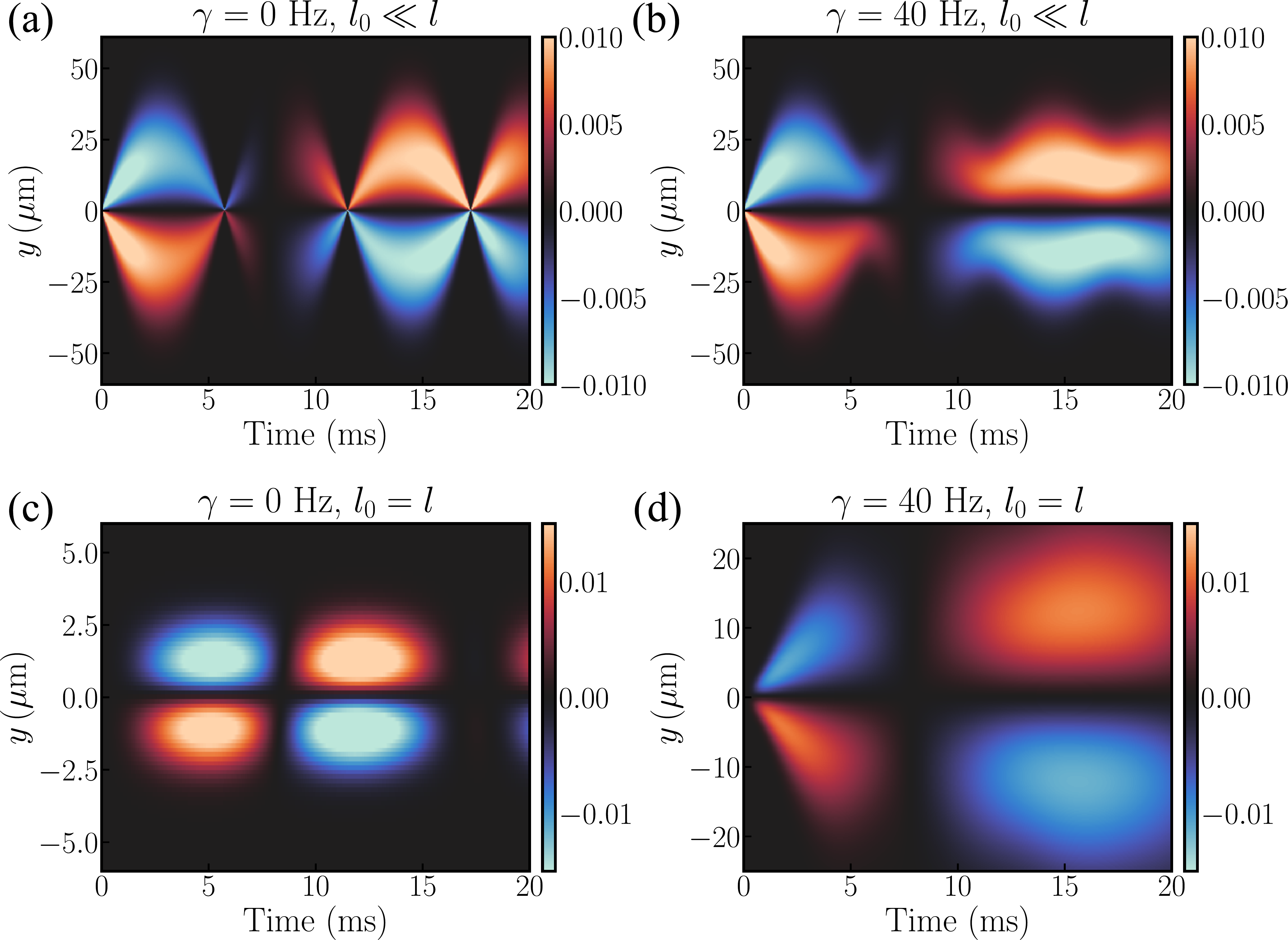}
  \caption{The same as in Fig.~\ref{fig:fig2} but with an additional magnetic field along the $y$ direction. The amplitude of the magnetic field is $\mu_B B/\hbar=100\,\mathrm{Hz}$.}
  \label{fig:fig3}
\end{figure}

Finally, we remark that Eq.~\eqref{modified master equation} allows us to include Zeeman-type terms, which naturally appear in ultracold atoms with synthetic SOC~\cite{Dalibard2011,Galitski2013}. 
To this end, we add the term $\mu_B\mathbf{B}\cdot\boldsymbol{\sigma}$ to $H_S$. Its presence strongly modifies the spin dynamics because SOC cannot be gauged out.  
Theoretical analysis also becomes more involved, since we cannot solve the system analytically for all values of $\alpha$ and $B$.  Still, we obtain closed-form expressions using tools of perturbation theory for $\alpha\to0$ (see the Supplementary Material). The effect of the magnetic field is illustrated in Fig.~\ref{fig:fig3}. Initially, the dynamics with the magnetic field is similar to the dynamics presented in Fig.~\ref{fig:fig2}. However, at later times we observe spin precession possible only in the presence of both SOC and the magnetic field. Spin precession leads to an exchange of domains with positive and negative values of $\bar \sigma_y$, and can be used for engineering the spin structure.

{\it Conclusions.} We analyzed Brownian-type motion of a spin-orbit coupled impurity with the goal to develop a simple theoretical tool that can be used to propose and analyze cold-atom-based quantum simulators. We introduced a master equation suitable for the problem. We tested it and illustrated its usefulness by interpreting available experimental data without SOC~\cite{Catani2012}. Our results suggested that the impurity does not experience any mass renormalization during quench dynamics at experimentally accessible temperatures. 
Finally, we demonstrated that systems with SOC can be studied analytically, and calculated observables that measure changes in population of the hyperfine states of the impurity atom.

A comparison between results of our theoretical study and experimental data (when available) can be used to understand the limits of applicability of a set of assumptions standard for studies of Brownian motion, such as the Markov approximation.
In addition, our findings pave the way for studies of various condensed matter systems where SOC and dissipation play a role. For example, models of the chirality induced spin selectivity (CISS) effect~\cite{Naaman2019,Evers2022} are typically based upon non-unitary time evolution (e.g., due to dissipation) and SOC, see, e.g.,~\cite{Guo2012,Fransson2020,Liu2021,Volosniev2021,Fransson2021,Barroso2022}. These effects are included in our model, hence, it can be developed into a testbed for studying CISS~\footnote{This would require to consider higher spatial dimensions as our results suggest that relevant effects due to SOC can be gauged out in 1D.}.

\acknowledgements
We thank Rafael Barfknecht for help at the initial stages of this project; Fabian Brauneis for useful discussions; Miguel A. Garcia-March, Georgios Koutentakis, and Simeon Mistakidis for comments on the manuscript.  
M.L. acknowledges support
by the European Research Council (ERC) Starting Grant
No. 801770 (ANGULON).

\newpage
\widetext

\appendix

\section{Technical details for derivations of the master equation}
\label{appendix A}

\subsection{Preliminaries}
We start by writing $H_B$ and $H_C$ as 
\begin{equation}
H_B  =\sum_{j=1}^N\bigg[\frac{\mathbf{p}_j^2}{2m_j} 
+\frac{1}{2}m_j\omega_j^2\mathbf{x}_j^2\bigg] = \sum_{j=1}^{N}\hbar\omega_j 
\mathbf{b}_j^{\dagger}
\mathbf{b}_j,
\label{hamiltonian harmonic bath}
\end{equation}
and
\begin{equation}
H_C  = -\mathbf{q} \cdot \sum_{j=1}^N c_j\mathbf{x}_j =
-\mathbf{q} \cdot \sum_{j=1}^N c_j\sqrt{\frac{\hbar}{2m_j\omega_j}}\big(
\mathbf{b}_j^{\dagger} + \mathbf{b}_j),
\label{coupling hamiltonian}
\end{equation}
where $\mathbf{b}^{\dagger}_j$ and $\mathbf{b}_j$ are the standard bosonic creation and annihilation 
operators, respectively.
For convenience, we define the operator
\begin{equation}
\mathcal{B} = \sum_{j=1}^N c_j 
\sqrt{\frac{\hbar}{2m_j\omega_j}}\big(\mathbf{b}_j^{\dagger} 
+ \mathbf{b}_j\big)\;.
\end{equation}
To derive the Born-Markov equation of motion for the impurity, we work in the interaction representation, where $\mathbf{q}$ and $\mathcal{B}$ are written as
\begin{equation}
\begin{aligned}
\mathbf{q}(t) & = e^{iH_S t/\hbar}\, \mathbf{q}\, e^{-iH_S\, t/\hbar}, \\
 & \\
\mathcal{B}(t) & = e^{iH_B t/\hbar}\, \mathcal{B}\, e^{-iH_B\, t/\hbar} = 
\sum_{j=1}^N c_j \sqrt{\frac{\hbar}{2m_j \omega_j}} \big(\mathbf{b}_je^{-i\omega_j t}
+ \mathbf{b}^{\dagger}_je^{i\omega_jt}
\big)\;. \\
\end{aligned}
\label{interaction picture q and B}
\end{equation} Here and in what follows, we write explicitly the time dependence (e.g., $\mathcal{B}\to\mathcal{B}(t)$) to designate the operators in the interaction representation.
The equation of motion for $\rho_{\text{tot}}$, namely $i\hbar \dot{\rho}_{\text{tot}} = [H_{\text{tot}},\rho_{\text{tot}}]$, 
is formally solved (in the interaction representation) by
\begin{equation}
\rho_{\text{tot}}(t) = \rho_{\text{tot}}(0) - \frac{i}{\hbar}\int_0^{t}dt' \big[
H_C(t'), \rho_{\text{tot}}(t')
\big]\;.
\label{integral form first step}
\end{equation}
Equation~\eqref{integral form first step} 
can be iterated one more time, leading to
\begin{equation}
\rho_{\text{tot}}(t) = \rho_{\text{tot}}(0) 
- \frac{i}{\hbar}\int_0^{t}dt' \big[
H_C(t'), \rho_{\text{tot}}(0)\big]
- \frac{1}{\hbar^2} \int_0^t dt' \int_0^{t'} dt'' \bigg[
H_C(t'),[H_C(t''),\rho_{\text{tot}}(t'')]
\bigg] \;.
\label{integral form second step}
\end{equation}
A time derivative of this equation reads as
\begin{equation}
\frac{d\rho_{\text{tot}}(t)}{dt} = -\frac{i}{\hbar}\big[
H_C(t),\rho_{\text{tot}}(0)
\big] - \frac{1}{\hbar^2}
\int_0^{t} dt' \bigg[
H_C(t), \big[
H_C(t'), \rho_{\text{tot}}(t')
\big]
\bigg]\;.
\label{master equation very general appendix}
\end{equation}

\subsection{Initial conditions}

Let us specify the initial conditions for Eq.~(\ref{master equation very general appendix}). For quench dynamics considered in the main text, we assume that the system
is put in contact with the environment at $t = 0$ such that
\begin{equation}
\rho_{\text{tot}}(0) = \rho_S(0) \otimes \rho_B(0),
\label{initial condition}
\end{equation}
where $\rho_B(0)$ is the equilibrium distribution for the bath, and $\rho_S(0)$ is the density matrix of the impurity. Under this condition,
the first term on the right-hand-side of Eq.~\eqref{master equation very general appendix} vanishes
when we trace over the degrees of freedom of the bath: 
\begin{equation}
\begin{aligned}
\text{Tr}_B \big[
H_C, \rho_S(0) \otimes \rho_B(0) 
\big] & = \text{Tr}_B \big[ - \mathbf{q}\,\mathcal{B}, \rho_S(0) \otimes \rho_B(0)\big] \\
& = \big[-\mathbf{q},\rho_S(0)\big] \text{Tr}_B \big[
\mathcal{B}, \rho_B(0)
\big] \\
& = \big[-\mathbf{q},\rho_S(0)\big] \langle\mathcal{B}\rangle_{\text{bath}} \\
& = 0\;.
\end{aligned}
\label{partial trace appendix}
\end{equation}
By taking the partial trace, $\mathrm{Tr}_B$, over Eq.~\ref{master equation very general appendix},
we derive
\begin{equation}
\frac{d\rho_S(t)}{dt} = - \frac{1}{\hbar^2}\int_0^{t} dt' \;\text{Tr}_B \bigg[
H_C(t),\big[
H_C(t'),\rho_{\text{tot}}(t')
\big]
\bigg]\;.
\label{master equation before born markov}
\end{equation}

\subsection{The Born-Markov approximation}

To deal with $\rho_{\text{tot}}(t)$ in Eq.~(\ref{master equation before born markov}), we employ the Born approximation, which assumes that the system-environment coupling is so 
weak that the bath is (almost) not affected by the dynamical evolution of the system. Technically speaking, it means
\begin{equation}
\rho_{\text{tot}}(t) \simeq \rho_S(t) \otimes \rho_B^{\text{eq}}\;,
\label{born}
\end{equation}
with $\rho_B^{\text{eq}}$ being the bath density matrix at thermal equilibrium. 

To further simplify the model, we adopt the coarse-grained
perspective on the time axis. In this Markovian picture, the decay of bath correlations
provides the shortest timescale ($\tau_B$); we cannot
resolve dynamical features with a comparable characteristic time. 
This means that we  replace
$t' \rightarrow t - t'$ and then let $\int_0^{t}dt' \rightarrow \int_0^{+\infty} 
dt'$. With these assumptions, Eq.~\eqref{master equation before born markov} leads to
\begin{equation}
\frac{d\rho_S(t)}{dt} = - \frac{1}{\hbar^2}\int_0^{+\infty} dt' \text{Tr}_B \bigg[
H_C(t),\big[
H_C(t-t'),\rho_S(t)\otimes
\rho_B^{\text{eq}}\big]
\bigg]\;.
\label{born markov master equation interaction supplementary}
\end{equation}
In the Schr{\"o}dinger representation, the Born-Markov master
equation is written down as
\begin{equation}
\begin{aligned}
\frac{d\rho_S}{dt} & = - \frac{i}{\hbar}\big[H_S,\rho_S\big] -
\frac{1}{\hbar^2} \int_0^{+\infty} dt'\; \text{Tr}_B \bigg[H_C,\big[H_C(-t'),
\rho_S(t)\otimes \rho_B^{\text{eq}}\big]\bigg] \\
& = - \frac{i}{\hbar}\big[H_S,\rho_S\big] 
- \frac{1}{\hbar^2} \int_0^{+\infty} dt'\, \mathcal{C}(t')\bigg[
\mathbf{q},\big[ \mathbf{Q}(-t'),\rho_S\big]
\bigg] 
+ \frac{i}{\hbar^2}\int_0^{+\infty} dt'\, \chi(t')\bigg[\mathbf{q},\lbrace \mathbf{Q}(-t'), 
\rho_S\rbrace\bigg]\;,
\end{aligned}
\label{born-markov master equation supplementary}
\end{equation}
where $\mathbf{Q}(t)$ is defined according to Eq. \eqref{position operator interaction}
in the main text.

\subsection{Autocorrelation functions}

The key quantities in Eq.~(\ref{born-markov master equation supplementary}) are the bath autocorrelation functions $\mathcal{C}(t)$ and
$\chi(t)$ defined as
\begin{equation}
\begin{aligned}
\mathcal{C}(t) & = \frac{1}{2} \sum_{j=1}^N c_j^2 \langle\lbrace \mathbf{x}_j(t),
	 \mathbf{x}_j(0)\rbrace\rangle_B   = 
 \hbar \int_0^{+\infty} d\omega\,J(\omega) \coth\bigg(\frac{\beta\hbar\omega}{2}\bigg) \cos(\omega t),\\
\chi(t) & = \frac{i}{2} \sum_{j=1}^N c_j^2 
\langle\big[\mathbf{x}_j(t),\mathbf{x}_j(0)\big]\rangle_B  = 
\hbar\int_0^{+\infty} d\omega\, J(\omega) \sin(\omega t)\;;
\end{aligned}
\end{equation}
they are 
related to noise and dissipation, respectively. The relation becomes clear by looking at 
Eq.~\eqref{born-markov master equation supplementary} and by considering the 
corrections beyond the (closed-system) Schr{\"o}dinger dynamics. These corrections are encoded
in a super-operator whose real part is proportional to 
$\big[
\mathbf{q},\big[ \mathbf{Q}(-t),\rho_S\big]
\big]$ with coefficient $\mathcal{C}(t)$; the imaginary part
is determined to $\chi(t)\big[\mathbf{q},\lbrace \mathbf{Q}(-t), 
\rho_S\rbrace\big]$.

The spectral function, $J(\omega)$, that enters in the autocorrelation functions is formally defined as
\begin{equation}
J(\omega) = \sum_{j=1}^N \frac{c_j^2}{2m_j \omega_j} \delta (\omega-\omega_j) \;.
\end{equation}
However, as mentioned in the main text, it is worth moving to the frequency continuum
and devise a reasonable, phenomenological choice for $J(\omega)$. One possibility
(employed in our work) is
\begin{equation}
J(\omega) = \frac{2m\gamma}{\pi} \, \omega \, \frac{\Omega_c^2}{\Omega_c^2 + \omega^2}\;.
\label{eq:Jomega_SM}
\end{equation}
It leads to the standard Ohmic dissipation ($J(\omega) \sim \omega$) at low frequencies.  Equation~(\ref{eq:Jomega_SM}) does not have any abnormal behavior (i.e., $J(\omega)$ does not diverge) at high frequencies thanks to the high-frequency cutoff determined by
the phenomenological parameter $\Omega_c$. This leads to meaningful theoretical results. Using the autocorrelation functions in 
Eq.~\eqref{born-markov master equation supplementary}, we derive 
Eq.~\eqref{modified master equation} of the main text.

\subsection{Calculation of integrals} 

Below we give an 
example of how the frequency integrals leading to 
Eq. \eqref{modified master equation} are computed.
By recalling the definition of $\mathbf{Q}(t)$ in Eq. \eqref{position operator interaction}
of the main text, we typically have to deal with 
\begin{equation}
- \frac{i}{2m\hbar^2} \int_0^{+\infty} dt'\,t'\chi(t') \big[
\mathbf{q},\lbrace \mathbf{p}/m + \mathbf{v}_{\text{SO}},\rho_S
\rbrace
\big]	 = 
- \frac{i}{\hbar m} \int_0^{+\infty} d\omega\, J(\omega) \int_0^{+\infty} dt'\,
t' \sin(\omega t') \big[\mathbf{q},\lbrace \mathbf{p}/m + \mathbf{v}_{\text{SO}},\rho_S
\rbrace
\big] \;.
\label{deriving usual damping}
\end{equation}
Now, we notice that the integral over time can be interpreted as a derivative of the
Dirac delta function. More precisely,
\begin{equation}
\begin{aligned}
\int_0^{+\infty} dt \, t \sin(\omega t)  = \frac{1}{2}\int_{-\infty}^{+\infty} dt \, t \sin(\omega t) & = 
- \pi \frac{\partial}{\partial \omega} \int_{-\infty}^{+\infty} \frac{dt}{2\pi}\cos(\omega t) 
 = - \pi \delta'(\omega)\;.
\end{aligned}
\end{equation}
Therefore, we can write Eq. \eqref{deriving usual damping} as
\begin{equation}
\begin{aligned}
- \frac{i}{\hbar m} \int_0^{+\infty} d\omega\, J(\omega) \int_0^{+\infty} dt'\,
t' \sin(\omega t') \big[\mathbf{q},\lbrace \mathbf{p}/m + \mathbf{v}_{\text{SO}},\rho_S
\rbrace
\big] & = - \frac{i\pi}{2\hbar m} \int_{-\infty}^{+\infty} d\omega \, \delta(\omega) \partial_{\omega}J(\omega)
\big[\mathbf{q},\lbrace \mathbf{p}/m + \mathbf{v}_{\text{SO}},\rho_S
\rbrace
\big] \\
& = - \frac{i\pi}{2\hbar m}\bigg[\lim_{\omega \rightarrow 0^{+}} J'(\omega) \bigg]
\big[\mathbf{q},\lbrace \mathbf{p}/m + \mathbf{v}_{\text{SO}},\rho_S
\rbrace
\big] \\
& = - \frac{i\gamma}{\hbar} \big[\mathbf{q},\lbrace \mathbf{p}/m + \mathbf{v}_{\text{SO}},\rho_S
\rbrace
\big]
\end{aligned}
\label{deriving usual damping 2}
\end{equation}

\subsection{Spin-orbit coupling}

Depending on the form of the spin-orbit coupling, the velocity 
operator, defined as $\mathbf{v}_{\mathrm{SO}}= \partial_t \mathbf{Q}$, with $\mathbf{Q}$ as in 
Eq. \eqref{position operator interaction} in the main text, is given by
\begin{equation}
v_{\mathrm{SO},1D} = \frac{p}{m} + \alpha\,s
\label{velocity 1d}
\end{equation}
for a strictly one-dimensional setup ($V_{SO} = s \alpha p$, $p$ being the particle momentum operator), or
\begin{equation}
\mathbf{v}_{\mathrm{SO}, 2D} = \frac{\mathbf{p}}{m}\, \sigma_0 - \frac{\alpha}{\hbar} \sigma_y\,
\mathbf{e}_x + \frac{\alpha}{\hbar} \sigma_x\,\mathbf{e}_y \;,
\label{velocity 2d}
\end{equation}
for a two-dimensional Rashba-like coupling, namely $V_{SO} = \alpha\big(-\sigma_y p_x + \sigma_x p_y\big)/\hbar$. The first terms on the right-hand-sides of Eqs.~(\ref{velocity 1d}) and (\ref{velocity 2d}) represent the standard relation between the momentum and the velocity. The other terms enter due to the presence of spin-orbit coupling. 

\subsection{Master equation in coordinate space}

The general form of the master equation in presence of SOC is presented 
in the main text, see~Eq. \eqref{modified master equation}. For our calculations, we use this equation in the position-space representation where a one-dimensional setup is described by the equation
\begin{equation}
\begin{aligned}
\frac{d\rho}{dt}=&\bigg[\frac{i\hbar}{2m}\left(\partial^2_y-\partial^2_{y^\prime}\right)-\gamma\left(y-y^\prime\right)\cdot\left(\partial_y-\partial_{y^\prime}\right)  - \frac{2m\gamma}{\beta\hbar^2}(y-y')^2 -\frac{i}{\hbar}\big(\mathcal{V}_\mathrm{ext}\left(y\right)- \mathcal{V}_\mathrm{ext}\left(y'\right)\big) \bigg]\rho \\
&\qquad  -\alpha\left(\sigma_x\partial_y \rho+\partial_{y^\prime}\rho\sigma_x\right) - \frac{im\gamma \alpha}{\hbar} (y-y')\big(\sigma_x\rho + \rho\sigma_x \big)\\
&\qquad \qquad \frac{i\mu_B\mathbf{B}}{\hbar}\cdot\left(\boldsymbol{\sigma}\rho- \rho\boldsymbol{\sigma}\right)+\frac{\hbar^2\gamma\beta}{8m}\left(\partial^2_y+\partial^2_{y^\prime}+2\partial_y\cdot\partial_{y^\prime}\right)\rho\;,
\end{aligned}
\label{master position_1D}
\end{equation}
where 
$\rho \equiv \langle \mathbf{r} |\rho_S| \mathbf{r}'\rangle$. 

For the sake of completeness, we also present the result for
two-dimensional setups:
\begin{align}
\frac{d\rho}{dt}=&\left[\frac{i\hbar}{2m}\left(\nabla^2_\mathbf{r}-\nabla^2_{\mathbf{r}^\prime}\right)-\gamma\left(\mathbf{r}-\mathbf{r}^\prime\right)\cdot\left(\nabla_\mathbf{r}-\nabla_{\mathbf{r}^\prime}\right)-\right. \nonumber \left. \frac{2m\gamma}{\beta\hbar^2}\left(\mathbf{r}-\mathbf{r}^\prime\right)^2-\frac{i}{\hbar}\left(\mathcal{V}_\mathrm{ext}\left(\mathbf{r}\right)-\mathcal{V}_\mathrm{ext}\left(\mathbf{r}^\prime\right)\right)\right]\rho \nonumber \\
&\qquad \qquad-\alpha\left(\sigma_x\partial_y\rho+\partial_{y^\prime}\rho\sigma_x-\sigma_y\partial_x\rho-\partial_{x^\prime}\rho\sigma_y\right)-\frac{im\gamma\alpha}{\hbar} \left[\left(y-y^\prime\right)\left(\sigma_x\rho+\rho\sigma_x\right)-\left(x-x^\prime\right)\left(\sigma_y\rho+\rho\sigma_y\right)\right]\nonumber \\
&\qquad \qquad \qquad-\frac{i\mu_B\mathbf{B}}{\hbar}\cdot\left(\boldsymbol{\sigma}\rho-\rho\boldsymbol{\sigma}\right)+\frac{\hbar^2\gamma\beta}{8m}\left(\nabla^2_\mathbf{r}+\nabla^2_{\mathbf{r}^\prime}+2\nabla_\mathbf{r}\cdot\nabla_{\mathbf{r}^\prime}\right)\rho.
\label{master position_2D}
\end{align}

\section{Technical details for calculations with the master equation}
\label{appendix B}
\subsection{Solution of the 1D master equation with $\alpha=0$}
When $\alpha=0$ or $B=0$ the master equation \eqref{master position_1D} can be solved exactly. Let us analyze these limits before moving to the case when both terms are present. First, we shall concentrate on the system without SOC, i.e., $\alpha=0$ and $B\neq0$. We write the density matrix as $\rho=\rho_0\sigma_0+\rho_1\sigma_x+\rho_2\sigma_y+\rho_3\sigma_z$ and assume (without loss of generality) that $\mathbf{B}\parallel\mathbf{y}$. Converting from $\rho_1$ and $\rho_3$ into $\rho_\pm=\rho_1\pm i\rho_3$, we have the equations
\begin{align}
\frac{d\rho_j}{d t}=&\left[\frac{i\hbar}{2m}\left(\partial^2_y-\partial^2_{y^\prime}\right)-\gamma\left(y-y^\prime\right)\left(\partial_y-\partial_{y^\prime}\right)-\frac{2m\gamma}{\beta\hbar^2}\left(y-y^\prime\right)^2-\frac{i\hbar}{2ml^4}\left(y^2-y^{\prime 2}\right)-\frac{2is\mu_BB}{\hbar}+ \right. \nonumber \\
&\hspace{220pt}\left.+\frac{\hbar^2\gamma\beta}{8m}\left(\partial^2_y+\partial^2_{y^\prime}+2\partial_y\partial_{y^\prime}\right)\right]\rho_j,
\label{eqn:master1D}
\end{align}
where $s=\pm$ for $j=\pm$ and zero otherwise. The solution of these equations is obtained using the method of characteristics~\cite{roy1999exact}. To employ this method, we first transform into the center-of-mass and relative coordinates $R=\left(y+y^\prime\right)/2$ and $r=y^\prime-y$. Then, we perform Fourier transform with respect to $R$
\begin{equation}
\rho_{j}\left(R,r,t\right)=\frac{1}{\sqrt{2\pi}}\int_{-\infty}^{+\infty}dKe^{iKR}\rho_{j}\left(K,r,t\right).    
\end{equation}
According to the method of characteristics, the solution for $\rho_{j}\left(K,r,t\right)$ has the form
\begin{equation}
\rho_{j}\left(K,r,t\right)=\rho_{j}\left(K^\prime,r^\prime,0\right)e^{aZ\left(K,r,t\right)+bZ^\prime\left(K,r,t\right)-\frac{2is\mu_BBt}{\hbar}},
\label{eqn:DensityMatrix}
\end{equation}
where $\rho_{j}\left(K^\prime,r^\prime,0\right)$ is determined from the initial conditions. The other quantities that enter Eq.~(\ref{eqn:DensityMatrix})  are defined as
\begin{align}
\label{eqn:a}
a&=\frac{\gamma}{2\beta m\left(\gamma^2-\omega^2\right)}, \\
b&=\frac{\gamma\beta\hbar^2\omega^2}{32m\left(\gamma^2-\omega^2\right)}, \\
Z\left(K,r,t\right)&=\frac{1}{\gamma}\left(K-\frac{r}{\lambda_+}\right)\left(K-\frac{r}{\lambda_-}\right)\left(1-e^{-2\gamma t}\right)-
\frac{m\lambda_+}{2\hbar}\left(K-\frac{r}{\lambda_+}\right)^2\left(1-e^{-\frac{2\hbar t}{m\lambda_+}}\right)-\nonumber \\
&\hspace{120pt}\frac{m\lambda_-}{2\hbar}\left(K-\frac{r}{\lambda_-}\right)^2\left(1-e^{-\frac{2\hbar t}{m\lambda_-}}\right), \\
Z^\prime\left(K,r,t\right)&=\frac{1}{\gamma}\left(K-\frac{r}{\lambda_+}\right)\left(K-\frac{r}{\lambda_-}\right)\left(1-e^{-2\gamma t}\right)-
\frac{m^3\omega^2\lambda^3_+}{2\hbar^3}\left(K-\frac{r}{\lambda_+}\right)^2\left(1-e^{-\frac{2\hbar t}{m\lambda_+}}\right)-\nonumber \\
&\hspace{120pt}\frac{m^3\omega^2\lambda^3_-}{2\hbar^3}\left(K-\frac{r}{\lambda_-}\right)^2\left(1-e^{-\frac{2\hbar t}{m\lambda_-}}\right), \\
\lambda_\pm&=\frac{\hbar}{m\omega^2}\left(\gamma\pm\sqrt{\gamma^2-\omega^2}\right), \\
K^\prime&=\frac{\left(\lambda_+ K-r\right)e^{-\frac{\hbar t}{m\lambda_+}}-\left(\lambda_- K-r\right)e^{-\frac{\hbar t}{m\lambda_-}}}{\lambda_+-\lambda_-}, \\
r^\prime&=\frac{\lambda_-\left(\lambda_+ K-r\right)e^{-\frac{\hbar t}{m\lambda_+}}-\lambda_+\left(\lambda_- K-r\right)e^{-\frac{\hbar t}{m\lambda_-}}}{\lambda_+-\lambda_-}.
\label{ZZp definition}
\end{align}
Notice that for $B=0$ this expression correspond to the known solution of the Caldeira-Legett equation~\cite{roy1999exact}.

The initial condition \eqref{eqn:InitialSpatial} reads in $K^\prime$ and $r^\prime$ variables as follows
\begin{equation}
\rho_{0}\left(K^\prime,r^\prime,0\right)=\frac{1}{2\sqrt{2\pi}}e^{-\frac{K^{\prime2}l_0^2}{4}-\frac{r^{\prime2}}{4l_0^2}-iK^\prime y_0};
\end{equation}
$\rho_{3}\left(K^\prime,r^\prime,0\right)=p\rho_{0}\left(K^\prime,r^\prime,0\right)$ and $p=1$ for spin polarized case and zero otherwise. For the sake of discussion, here we have also included $y_0$ -- the shift of the initial wave packet.  The density  matrix in real space is cumbersome. We do not present it here since in this work we are not interested in spatial correlations. Instead, we focus on local observables for which $y=y^\prime$, $R=y$ and $r=0$. To calculate them, we notice that $K^\prime=g_1(t)K$, $r^\prime=g_2(t)K$, $Z(K,0,t) =g_3(t)K^2$, $Z^\prime(K,0,t)=g_4(t)K^2$,
where functions $g_i(t)$ are defined as
\begin{align}
g_1\left(t\right)&=\frac{\lambda_+e^{-\frac{\hbar t}{m\lambda_+}}-\lambda_-e^{-\frac{\hbar t}{m\lambda_-}}}{\lambda_+-\lambda_-}, \\ 
g_2\left(t\right)&=\frac{\lambda_+\lambda_-}{\lambda_+-\lambda_-}\left(e^{-\frac{\hbar t}{m\lambda_+}}-e^{-\frac{\hbar t}{m\lambda_-}}\right), \\
g_3\left(t\right)&=\frac{1}{\gamma}\left(1-e^{-2\gamma t}\right)-
\frac{m\lambda_+}{2\hbar}\left(1-e^{-\frac{2\hbar t}{m\lambda_+}}\right)-\frac{m\lambda_-}{2\hbar}\left(1-e^{-\frac{2\hbar t}{m\lambda_-}}\right),\\
g_4\left(t\right)&=\frac{1}{\gamma}\left(1-e^{-2\gamma t}\right)-
\frac{m^3\omega^2\lambda^3_+}{2\hbar^3}\left(1-e^{-\frac{2\hbar t}{m\lambda_+}}\right)-\frac{m^3\omega^2\lambda^3_-}{2\hbar^3}\left(1-e^{-\frac{2\hbar t}{m\lambda_-}}\right).
\end{align}
Now, we can compute the densities from Eq.~(\ref{eqn:DensityMatrix})
\begin{align}
\rho_0(y,y,t)&=\frac{1}{4\sqrt{\pi b_0(t)}}e^{-\frac{\left(y-y_0g_1(t)\right)^2}{4b_0(t)}},\\
\rho_1(y,y,t)&=p\rho_0(y,y,t)\sin\left(\frac{2\mu B t}{\hbar}\right), \\
\rho_2(y,y,t)&=0, \\
\rho_3(y,y,t)&=p\rho_0(y,y,t)\cos\left(\frac{2\mu B t}{\hbar}\right),
\end{align}
where $b_0(t)=g^2_1(t)l^2_0/4+g^2_2(t)/4l^2_0-ag_3(t)-bg_4(t)$.  
The typical observables can also be easily calculated:
\begin{align}
\langle y\rangle&=y_0g_1\left(t\right), \\
\label{eq:analytic_exp}
\langle y^2\rangle&=y^2_0g^2_1(t)+2b_0(t),\\
\langle \sigma_x\rangle&=p\sin\left(\frac{2\mu_B B t}{\hbar}\right), \\
\langle y\sigma_x\rangle&=py_0g_1\left(t\right)\sin\left(\frac{2\mu_B B t}{\hbar}\right), \\
\langle \sigma_y\rangle&=0, \\
\langle y\sigma_y\rangle&=0,\\
\langle\sigma_z\rangle&=p\cos\left(\frac{2\mu_B B t}{\hbar}\right), \\
\langle y\sigma_z\rangle&=py_0g_1\left(t\right)\cos\left(\frac{2\mu_B B t}{\hbar}\right).
\end{align}

\subsection{Comparison with the fit used in the experiment}
In this section we compare the result of Eq.~(\ref{eq:analytic_exp}) with Eq.~(\ref{eq:fit_experiment}) of the main text, which we repeat here for convenience
\begin{equation}
    \bar y=\bar y_0+\mathcal{A}_1 t - \mathcal{A}_2 e^{-\Gamma \Omega t}\cos[\sqrt{1-\Gamma^2}\Omega(t-t_0)],
    \label{eq:fit_experiment_sup}
\end{equation}
where $\bar y=\sqrt{\langle y^2\rangle}$.
The functions $g_i(t)$ that enter Eq.~(\ref{eq:analytic_exp}) have the form
\begin{align}
g_1(t)&=\frac{\gamma e^{-\gamma t}}{\sqrt{\omega^2-\gamma^2}}\sin\left(\sqrt{\omega^2-\gamma^2}t\right)+e^{-\gamma t} \cos\left(\sqrt{\omega^2-\gamma^2}t\right), \\
g_2(t)&=\frac{\hbar e^{-\gamma t}}{m\sqrt{\omega^2-\gamma^2}}\sin\left(\sqrt{\omega^2-\gamma^2}t\right), \\
g_3(t)&=\frac{\omega^2-\gamma^2}{\gamma\omega^2}-\frac{1}{\gamma}e^{-2\gamma t}-\frac{\sqrt{\omega^2-\gamma^2} e^{-2\gamma t}}{\omega^2}\sin\left(2\sqrt{\omega^2-\gamma^2}t\right)+\frac{\gamma e^{-2\gamma t}}{\omega^2}\cos\left(2\sqrt{\omega^2-\gamma^2}t\right), \\
g_4(t)&=\frac{\omega^4-4\gamma^4+3\gamma^2\omega^2}{\gamma\omega^4}-\frac{1}{\gamma}e^{-2\gamma t}-\frac{\left(4\gamma^2-\omega^2\right)\sqrt{\omega^2-\gamma^2} e^{-2\gamma t}}{\omega^4}\sin\left(2\sqrt{\omega^2-\gamma^2}t\right)+\nonumber \\ &\hspace{200pt}\frac{\left(4\gamma^3-3\gamma\omega^2\right) e^{-2\gamma t}}{\omega^4}\cos\left(2\sqrt{\omega^2-\gamma^2}t\right).
\end{align}
In the experiment $y_0=0$, $\gamma\ll \omega$ and $k_BT\gg\hbar\omega$, thus, we can ignore the effect of the minimally invasive term, and approximate $\langle y^2\rangle$ as
\begin{equation}
\langle y^2\rangle=2b_0(t)\approx\frac{1}{2\beta m\omega^2}+\left(\frac{l^4_0+l^4}{4l^2_0}-\frac{1}{2\beta m\omega^2}\right)e^{-2\gamma t}+\frac{l^4_0-l^4}{4l^2_0}e^{-2\gamma t}\cos\left(2\omega t\sqrt{1-\left(\frac{\gamma}{\omega}\right)^2}\right).
\label{eq:analytic_simpl}
\end{equation}
Note that in the experiment $l_0$ is the smallest length scale, i.e, $l_0\ll l$, and that $l^4/4l^2_0$ is comparable to $1/2\beta m\omega^2$. Therefore, at long times ($t\gtrsim 1/\gamma$) we can estimate $\langle y^2\rangle$
by disregarding the exponentially decaying second term in (\ref{eq:analytic_simpl}). After these simplifications, we derive
\begin{equation}
\langle y^2\rangle\approx\frac{l^2}{2\beta \hbar\omega}-\frac{l^4}{4l^2_0}e^{-2\gamma t}\cos\left(2\omega t\sqrt{1-\left(\frac{\gamma}{\omega}\right)^2}\right).
\label{eq:analytic_simpl_2}
\end{equation}
Comparing (\ref{eq:analytic_simpl_2}) with (\ref{eq:fit_experiment_sup}), we can identify $\Omega=2\omega$, $\Gamma=\gamma/\omega$, $\bar y_0=l/\sqrt{2\beta\hbar\omega}$,  $\mathcal{A}_2=l^4/8\bar y_0 l^2_0$. We see that the dynamics of the system is fully determined by $l_0$ and $\gamma$. Assuming that $\gamma$ is measured in the experiment, the only free parameter is the initial energy fixed by $l_0$. Note that the width of the steady state does not depend on the initial state, in agreement with general postulates of thermodynamics. 

The values of $\mathcal{A}_1$ and $t_0$ determine, in particular, the initial `inflation', i.e., increase of the oscillation amplitude. These parameters describe phenomenologically the effect of exponentially decaying terms. Since they are strongly model-dependent (i.e., depend on the choice of the fitting function), we do not discuss them here.

\subsection{Solution of the 1D master equation with $B=0$}
Here, we consider systems with $\alpha\neq0$ and $B=0$. As was noted in the main text, in this case SOC can be gauged out by transformation
\begin{equation}
\rho\left(y,y^\prime,t\right)=e^{-\frac{i m\alpha \sigma_x y}{\hbar}}f\left(y,y^\prime,t\right)e^{\frac{i m\alpha \sigma_x y^\prime}{\hbar}},
\label{eqn:gaugeAppendix}
\end{equation}
where $f_S\left(y,y^\prime,t\right)$ satisfies Eq.~\eqref{eqn:master1D} with $B=0$; the function $f$ has the form presented in Eq.~\eqref{eqn:DensityMatrix}.  Presence of SOC modifies the initial condition for $f_{j}\left(K^\prime, r^\prime,0\right)$:
\begin{align}
\label{eqn:initf}
f_{0}\left(K^\prime, r^\prime,0\right)&=\cos\left(\frac{m\alpha r^\prime}{\hbar}\right)\rho_{0}\left(K^\prime,r^\prime,0\right) \\
f_{1}\left(K^\prime, r^\prime,0\right)&=-i\sin\left(\frac{m\alpha r^\prime}{\hbar}\right)\rho_{0}\left(K^\prime,r^\prime,0\right) \\
f_{2}\left(K^\prime, r^\prime,0\right)&=\frac{ip}{2}\left(e^{-\frac{m\alpha K^\prime l_0^2+2im\alpha y_0}{\hbar}}-e^{\frac{m\alpha K^\prime l_0^2+2im\alpha y_0}{\hbar}}\right)e^{\frac{m^2\alpha^2l_0^2}{\hbar^2}}\rho_{0}\left(K^\prime,r^\prime,0\right) \\
f_{3}\left(K^\prime, r^\prime,0\right)&=\frac{p}{2}\left(e^{-\frac{m\alpha K^\prime l_0^2+2im\alpha y_0}{\hbar}}+e^{\frac{m\alpha K^\prime l_0^2+2im\alpha y_0}{\hbar}}\right)e^{\frac{m^2\alpha^2l_0^2}{\hbar^2}}\rho_{0}\left(K^\prime,r^\prime,0\right).
\label{eqn:initl}
\end{align}
From these density matrices, we can derive the densities via inverse Fourier tranform 
\begin{align}
\label{eqn:diagf_0}
f_0(y,y,t)&=\frac{1}{8\sqrt{\pi b_0(t)}}\left(e^{-\frac{\left(y-y_0g_1(t)-\frac{m\alpha g_2(t)}{\hbar}\right)^2}{4b_0(t)}}+e^{-\frac{\left(y-y_0g_1(t)+\frac{m\alpha g_2(t)}{\hbar}\right)^2}{4b_0(t)}}\right),\\
f_1(y,y,t)&=\frac{1}{8\sqrt{\pi b_0(t)}}\left(e^{-\frac{\left(y-y_0g_1(t)-\frac{m\alpha g_2(t)}{\hbar}\right)^2}{4b_0(t)}}-e^{-\frac{\left(y-y_0g_1(t)+\frac{m\alpha g_2(t)}{\hbar}\right)^2}{4b_0(t)}}\right), \\
f_2(y,y,t)&=-\frac{ip e^{\frac{ m^2\alpha^2 l^2_0}{\hbar^2}}}{8\sqrt{\pi b_0(t)}}\left(e^{\frac{2i m\alpha y_0}{\hbar}}e^{-\frac{\left(y-y_0g_1(t)-\frac{i m\alpha g_1(t) l^2_0}{\hbar}\right)^2}{4b_0(t)}}-e^{-\frac{2i m\alpha y_0}{\hbar}}e^{-\frac{\left(y-y_0g_1(t)+\frac{i m\alpha g_1(t) l^2_0}{\hbar}\right)^2}{4b_0(t)}}\right), \\
f_3(y,y,t)&=\frac{p e^{\frac{ m^2\alpha^2 l^2_0}{\hbar^2}}}{8\sqrt{\pi b_0(t)}}\left(e^{\frac{2i m\alpha y_0}{\hbar}}e^{-\frac{\left(y-y_0g_1(t)-\frac{i m\alpha g_1(t) l^2_0}{\hbar}\right)^2}{4b_0(t)}}+e^{-\frac{2i m\alpha y_0}{\hbar}}e^{-\frac{\left(y-y_0g_1(t)+\frac{i m\alpha  g_1(t) l^2_0}{\hbar}\right)^2}{4b_0(t)}}\right).
\label{eqn:diagf_3}
\end{align}
After straightforward but tedious calculations, we derive time dynamics of observables
\begin{align}
\langle y\rangle&=y_0g_1\left(t\right), \\
\langle y^2\rangle&=y^2_0g^2_1(t)+2b_0(t)+\frac{m^2\alpha^2g^2_2(t)}{\hbar^2}, \\
\langle \sigma_x\rangle&=0, \\
\langle y\sigma_x\rangle&=\frac{m\alpha}{\hbar}g_2\left(t\right), \\
\langle \sigma_y\rangle&=pe^{\frac{m^2\alpha^2}{\hbar^2}\left(l^2_0\left(1+2g_1(t)\right)-4b_0(t)\right)}\sin\left(\frac{2m\alpha y_0}{\hbar}\left(1-g_1(t)\right)\right), \\
\langle y\sigma_y\rangle&=pe^{\frac{m^2\alpha^2}{\hbar^2}\left(l^2_0\left(1+2g_1(t)\right)-4b_0(t)\right)}\left[\frac{m\alpha}{\hbar}\left(g_1(t)l^2_0-4b_0(t)\right)\cos\left(\frac{2m\alpha y_0}{\hbar}\left(1-g_1(t)\right)\right)-\right. \nonumber \\
&\hspace{180pt}\left.y_0g_1(t)\sin\left(\frac{2m\alpha y_0}{\hbar}\left(1-g_1(t)\right)\right)\right], \label{eq:app_ysigmaz}\\
\langle \sigma_z\rangle&=pe^{\frac{m^2\alpha^2}{\hbar^2}\left(l^2_0\left(1+2g_1(t)\right)-4b_0(t)\right)}\cos\left(\frac{2m\alpha y_0}{\hbar}\left(1-g_1(t)\right)\right),\\
\langle y\sigma_z\rangle&=pe^{\frac{m^2\alpha^2}{\hbar^2}\left(l^2_0\left(1+2g_1(t)\right)-4b_0(t)\right)}\left[-\frac{m\alpha}{\hbar}\left(g_1(t)l^2_0-4b_0(t)\right)\sin\left(\frac{2m\alpha y_0}{\hbar}\left(1-g_1(t)\right)\right)+\right. \nonumber \\
&\hspace{180pt}\left.y_0g_1(t)\cos\left(\frac{2m\alpha y_0}{\hbar}\left(1-g_1(t)\right)\right)\right].    
\end{align}
We see that the presence of SOC modifies the time dynamics of spin-independent observables, such as $\langle y^2\rangle$. More importantly, it also generates spin dynamics, which was absent for $\alpha=0$.
To illustrate this dynamics, we consider $\langle y\sigma_y\rangle$ assuming that the initial state is spin-polarized in the $\mathbf{z}$-direction (i.e., $p=1$) and the initial packet is at the centre of the well (i.e., $y_0=0$). According to Eq.~(\ref{eq:app_ysigmaz}), SOC rotates the spin of the impurity, adding a component along the $\mathbf{y}$-direction. This effect can be understood already at the level of a one-body Schr{\"o}dinger equation. Presence of dissipation leads to a steady state for $t\gg1/\gamma$. Indeed, in this case $g_1\left(t\right)\rightarrow0$ and $g_2\left(t\right)\rightarrow0$, but $g_3\left(t\right)\rightarrow\left(\omega^2-\gamma^2\right)/\left(\omega^2\gamma\right)$ and  $g_4\left(t\right)\rightarrow\left(\omega^4+3\gamma^2\omega^2-4\gamma^4\right)/\left(\omega^4\gamma\right)$, leading to a finite value of $\langle y\sigma_y\rangle$ (see also the main text).

\subsection{Solution of the 1D master equation with $B\neq0$ and $\alpha\neq0$}
Finally, we consider the case with finite magnetic fields and SOC, i.e., $B\neq0$ and $\alpha\neq0$. We perform the gauge transformation presented in Eq.~\eqref{eqn:gaugeAppendix} and work with the function $f$. For the sake of discussion, we assume that the initial condition for $f\left(y,y^\prime,t\right)$ is as for the $B=0$ case; we also assume that $\mathbf{B}\parallel\mathbf{y}$. Assuming that SOC is weak, we restrict our calculations to the first order in $\alpha$.
The corresponding equations for $f$ [$f\left(y,y^\prime,t\right)=f_0\sigma_0+f_1\sigma_x+f_2\sigma_y+f_3\sigma_z$] are
\begin{align}
\frac{d f_0}{dt}&=\mathcal{L}f_0+\frac{2i\mu_B Bm\alpha}{\hbar^2}\left(y-y^\prime\right)f_3, \\
\frac{d f_1}{dt}&=\mathcal{L}f_1+\frac{2\mu_BB}{\hbar}f_3+\frac{2\mu_B Bm\alpha}{\hbar^2}\left(y+y^\prime\right)f_2, \label{eq:app:f1withBalpha}\\
\frac{d f_2}{dt}&=\mathcal{L}f_2-\frac{2\mu_B Bm\alpha}{\hbar^2}\left(y+y^\prime\right)f_1, \\
\frac{d f_3}{dt}&=\mathcal{L}f_3-\frac{2\mu_BB}{\hbar}f_1+\frac{2i\mu_B Bm\alpha}{\hbar^2}\left(y-y^\prime\right)f_0,
\end{align}
where $\mathcal{L}$ is the operator that reproduces the right-hand-side of the master equation with $B=0$ and $\alpha=0$. From the initial conditions \eqref{eqn:initf}-\eqref{eqn:initl}, it is clear that $f_2\propto\alpha$ even when $B=0$. Therefore, the last term of Eq.~(\ref{eq:app:f1withBalpha}) will be second order in $\alpha$ and can be ignored. To proceed, we use the expansion $f_i=f^0_i+\alpha f^1_i$, where $f^0_i$ satisfies the equations with $\alpha=0$ (cf.~Eq.~\eqref{eqn:DensityMatrix}).
Then, we solve the system of equations (it is convenient to solve for $f_\pm=f_1\pm i f_3$). 
For the densities in the leading order, we have
\begin{align}
f^0_0\left(y,y,t\right)&=f_0\left(y,y,t\right), \\
f^0_1\left(y,y,t\right)&=f_1\left(y,y,t\right)\cos\left(\frac{2\mu_B B t}{\hbar}\right)+f_3\left(y,y,t\right)\sin\left(\frac{2\mu_B B t}{\hbar}\right), \\
f^0_2\left(y,y,t\right)&=f_2\left(y,y,t\right), \\
f^0_3\left(y,y,t\right)&=-f_1\left(y,y,t\right)\sin\left(\frac{2\mu_B B t}{\hbar}\right)+f_3\left(y,y,t\right)\cos\left(\frac{2\mu_B B t}{\hbar}\right),
\end{align}
where on the right-hand-side we have density matrices from Eqs.~(\ref{eqn:diagf_0}-\ref{eqn:diagf_3}). The corresponding functions $f^1_i$ are
\begin{align}
f^1_{0}\left(y,y,t\right)&=\frac{2\mu_B B m}{\hbar^2}\mathrm{Im}\left[b_1(t)\frac{\partial}{\partial y}\left(f^0_1\left(y,y,t\right)-if^0_3\left(y,y,t\right)\right)\right], \\
f^1_{1}\left(y,y,t\right)&=\frac{2\mu_B B m}{\hbar^2}\mathrm{Im}\left[b_1(t)\frac{\partial}{\partial y}f^0_0\left(y,y,t\right)\right], \\
f^1_{2}\left(y,y,t\right)&=-\frac{4\mu_B B m}{\hbar^2}\mathrm{Re}\left[b_2(t)\frac{\partial}{\partial y}\left(f^0_1\left(y,y,t\right)-if^0_3\left(y,y,t\right)\right)\right] \nonumber  \\
&\hspace{120pt}+\frac{4\mu_B B m y_0}{\hbar^2}f^0_1\left(y,y,t\right), \\
f^1_{3}\left(y,y,t\right)&=-\frac{2\mu_B B m}{\hbar^2}\mathrm{Re}\left[b_1(t)\frac{\partial}{\partial y}f^0_0\left(y,y,t\right)\right],
\end{align}
where
\begin{align}
b_1(t)&=\frac{\frac{\hbar}{m}-\left(\frac{\hbar}{m}g_1\left(t\right)+\frac{2i\mu_B B}{\hbar}g_2\left(t\right)\right)e^{-\frac{2i\mu_B Bt}{\hbar}}}{\omega^2+\frac{4i\mu_B B\gamma}{\hbar}-\frac{4\mu^2_B B^2}{\hbar^2}}, \\
b_2(t)&=\left(\frac{2a\left(\omega^2-\gamma^2\right)}{\gamma\omega^2}+\frac{b\left(2\omega^4+6\gamma^2\omega^2-8\gamma^4\right)}{\gamma\omega^4}\right)\frac{\left(2\gamma+\frac{2i\mu_B B t}{\hbar}\right)\left(1-g_1(t)e^{-\frac{2i\mu_B B t}{\hbar}}\right) +\frac{m\omega^2}{\hbar}g_2(t)e^{-\frac{2i\mu_B B t}{\hbar}}}{\omega^2+\frac{4i\mu_BB\gamma}{\hbar}-\frac{4\mu^2_BB^2}{\hbar^2}}\nonumber \\
&-\frac{4mb\left(\omega^2-\gamma^2\right)}{\hbar\omega^2}b_1(t)+\left[\left(\frac{l^2_0}{2}+\frac{2a\left(\omega^2-\gamma^2\right)}{\gamma\omega^2}+\frac{b\left(2\omega^4+6\gamma^2\omega^2-8\gamma^4\right)}{\gamma\omega^4}\right)g_1(t)-\frac{4mb\left(\omega^2-\gamma^2\right)}{\hbar\omega^2}g_2(t)\right]g_5(t) \nonumber \\
&\hspace{120pt}+\left[\left(\frac{1}{2l^2_0}+\frac{2\left(a+b\right)m^2\left(\omega^2-\gamma^2\right)}{\hbar^2\gamma}\right)g_2(t)-\frac{4bm\left(\omega^2-\gamma^2\right)}{\hbar\omega^2}g_1(t)\right]g_6(t), \\
g_5(t)&=\frac{1}{\lambda_+-\lambda_-}\left(\frac{\lambda_+ \left(e^{-\frac{\hbar t}{m \lambda_+}}-e^{-\frac{2i\mu_B Bt}{\hbar}}\right)}{\frac{\hbar}{m\lambda_+}-\frac{2i\mu_B B}{\hbar}}-\frac{\lambda_- \left(e^{-\frac{\hbar t}{m \lambda_-}}-e^{-\frac{2i\mu_B Bt}{\hbar}}\right)}{\frac{\hbar}{m\lambda_-}-\frac{2i\mu_B B}{\hbar}}\right), \\
g_6(t)&=\frac{\lambda_+\lambda_-}{\lambda_+-\lambda_-}\left(\frac{\left(e^{-\frac{\hbar t}{m \lambda_+}}-e^{-\frac{2i\mu_B Bt}{\hbar}}\right)}{\frac{\hbar}{m\lambda_+}-\frac{2i\mu_B B}{\hbar}}-\frac{ \left(e^{-\frac{\hbar t}{m \lambda_-}}-e^{-\frac{2i\mu_B Bt}{\hbar}}\right)}{\frac{\hbar}{m\lambda_-}-\frac{2i\mu_B B}{\hbar}}\right)
\end{align}

The observables have the form
\begin{align}
\langle y\rangle&\approx y_0g_1\left(t\right)+\frac{2p\alpha\mu_B Bm}{\hbar^2}\mathrm{Re}\left[b_1(t)e^{\frac{2i\mu_B B t}{\hbar}}\right], \\
\langle y^2\rangle &\approx y^2_0g^2_1(t)+2b_0(t)+\frac{4p\alpha\mu_B Bm}{\hbar^2}y_0g_1(t)\mathrm{Re}\left[b_1(t)e^{\frac{2i\mu_B B t}{\hbar}}\right], \\
\langle \sigma_x\rangle&\approx p\sin\left(\frac{2\mu_B B t}{\hbar}\right), \\
\langle y\sigma_x\rangle&\approx\frac{m\alpha}{\hbar}g_2\left(t\right)\cos\left(\frac{2\mu_B B t}{\hbar}\right)+py_0g_1\left(t\right)\sin\left(\frac{2\mu_B B t}{\hbar}\right)-\frac{2\alpha\mu_B Bm}{\hbar^2}\mathrm{Im}\left[b_1(t)\right], \\
\langle \sigma_y\rangle &\approx \frac{2pm\alpha y_0}{\hbar}\left(1-g_1(t)\right), \\
\langle y\sigma_y\rangle&\approx\frac{pm\alpha}{\hbar}g_1\left(t\right)\left(l_0^2+2y^2_0\right)-\frac{pm\alpha}{\hbar}\left[g^2_1\left(t\right)\left(l_0^2+2y^2_0\right)+\frac{g^2_2\left(t\right)}{l_0^2}-4\left(ag_3\left(t\right)+bg_4\left(t\right)\right)\right]\times\nonumber \\
&\cos\left(\frac{2\mu_B B t}{\hbar}\right)+\frac{4p\alpha\mu_BBm}{\hbar^2}\mathrm{Im}\left[b_2e^{\frac{2i\mu_B B t}{\hbar}}\right]+\frac{4p\alpha\mu_B Bmy^2_0}{\hbar^2}g_1(t)\sin\left(\frac{2\mu_B B t}{\hbar}\right),\\
\langle \sigma_z\rangle&\approx p\cos\left(\frac{2\mu_B B t}{\hbar}\right), \\
\langle y\sigma_z\rangle&\approx py_0g_1\left(t\right)\cos\left(\frac{2\mu_B B t}{\hbar}\right)-\frac{m\alpha}{\hbar}g_2\left(t\right)\sin\left(\frac{2\mu_B B t}{\hbar}\right)+\frac{2\alpha\mu_B Bm}{\hbar^2}\mathrm{Re}\left[b_1(t)\right].    
\end{align}
These expressions are accurate for small values of $\alpha$, and off-resonant magnetic fields. [The resonance is located at $\hbar\omega=2\mu_B B$.]

\bibliography{apssamp}

\end{document}